\documentclass[runningheads]{llncs}
\usepackage[T1]{fontenc}
\usepackage{amsmath}
\usepackage{subfigure}
\usepackage{booktabs}

\usepackage{graphicx}
\begin{document}
\title{Towards Reliable Pediatric Brain Tumor Segmentation: Task-Specific nnU-Net Enhancements}
%
%
\author{Xiaolong Li\inst{1}\orcidID{0009-0000-3611-8885} \and Zhi-Qin John Xu\inst{1,*}\orcidID{0000-0003-0627-3520}\and
        Yan Ren\inst{2,*}\orcidID{0000-0001-5993-9248}
        \and
        Tianming Qiu\inst{3,*}\orcidID{0000-0003-1089-4717} \and
        Xiaowen Wang\inst{3,*}\orcidID{0009-0004-5559-0981}}

\authorrunning{Xiaolong Li et al.}
\titlerunning{Towards Reliable Pediatric Brain Tumor Segmentation}

\institute{ \textsuperscript{1} Institute of Natural Sciences, School of Mathematical Sciences, MOE-LSC, Shanghai Jiao Tong University, Shanghai, China\\ \email{15369855310@sjtu.edu.cn}\\
\email{xuzhiqin@sjtu.edu.cn}\\
\textsuperscript{2} Department of Radiology, Huashan Hospital, Fudan
University, Shanghai, China \\
\email{renyan\_richard@aliyun.com} \\
\textsuperscript{3} Department of Neurosurgery, Huashan Hospital, Fudan University, Shanghai, China \\
\email{tianming2100@126.com, apolloslisy@126.com} \\
} 
\maketitle 

\let\origfootnote\thefootnote
\renewcommand{\thefootnote}{}
\footnotetext{\textsuperscript{*} Corresponding authors.}
\let\thefootnote\origfootnote

\begin{abstract}

Accurate segmentation of pediatric brain tumors in multi-parametric magnetic resonance imaging (mpMRI) is critical for diagnosis, treatment planning, and monitoring, yet faces unique challenges due to limited data, high anatomical variability, and heterogeneous imaging across institutions. In this work, we present \textbf{an advanced nnU-Net framework} tailored for \textbf{BraTS 2025 Task-6 (PED)}, the largest public dataset of pre-treatment pediatric high-grade gliomas. Our contributions include: (1) a widened residual encoder with squeeze-and-excitation (SE) attention; (2) 3D depthwise separable convolutions; (3) a specificity-driven regularization term; and (4) small-scale Gaussian weight initialization. We further refine predictions with two postprocessing steps. Our models achieved first place on the Task-6 validation leaderboard, attaining lesion-wise Dice scores of \textbf{0.759 (CC)}, \textbf{0.967 (ED)}, \textbf{0.826 (ET)}, \textbf{0.910 (NET)}, \textbf{0.928 (TC)} and \textbf{0.928 (WT)}.

\end{abstract}

\section{Introduction}

\

\textbf{The Brain Tumor Segmentation (BraTS) Challenge}~\cite{karargyris2023federated}~\cite{kazerooni2024brain}~\cite{kazerooni2023brain} has served as a cornerstone in advancing automated neuro-oncological imaging analysis. By releasing large-scale, high-quality annotated datasets and formulating clinically relevant tasks, BraTS has driven innovation in algorithmic segmentation and classification of brain tumors. Continuing this mission, the BraTS 2025 Lighthouse Challenge introduces eleven diverse tasks targeting key translational gaps in brain tumor AI solutions, including segmentation, synthesis, and classification across different tumor types, age groups, and imaging conditions.

\begin{figure}[h]
    \centering
    \includegraphics[width=0.8\linewidth]{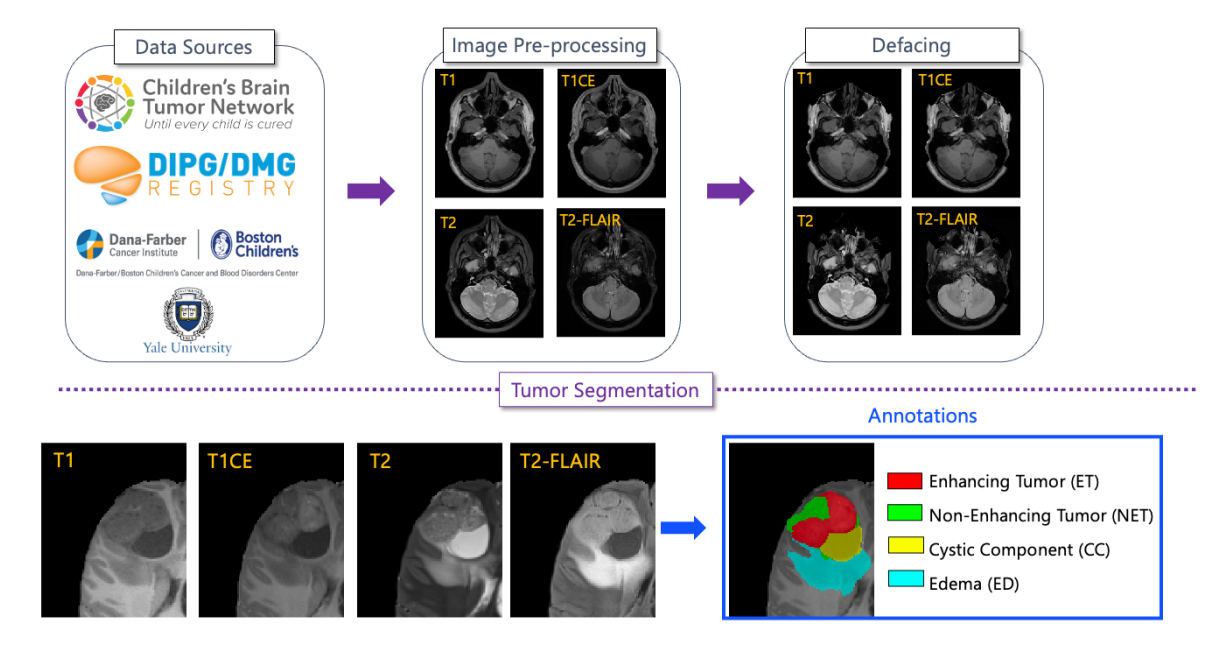}
    \caption{Graphical representation of data processing and annotations in pediatric brain tumors. Top panel presents the processing pipeline, and the bottom panel illustrates the annotated tumor subregions along with mpMRI structural scans (T1, T1CE, T2, and T2-FLAIR). Tumor subregions include the enhancing tumor (ET - red), non-enhancing tumor (NET - green), cystic component (CC -
yellow), and edema (ED - teal) regions.
}
    \label{PED}
\end{figure}

Task-6 (PED) of the BraTS 2025 Challenge focuses on a particularly underexplored and clinically significant domain: automatic segmentation of pre-treatment pediatric brain tumors. This task leverages the largest publicly available, expert-annotated cohort of high-grade pediatric brain tumors to date, aggregating multi-parametric MRI data~\ref{PED}~\cite{kazerooni2024brain}~\cite{kazerooni2023brain} from globally recognized pediatric oncology consortia.

In this work, we propose \textbf{An Advanced nnU-Net Framework for BraTS-2025 PED} to tackle the unique challenges of pediatric tumor segmentation.

The main contributions of this work are as follows:

\begin{itemize}

\item \textbf{Widened residual encoder with attention in nnU-Net~\cite{isensee2021nnu} architecture}.

\item \textbf{Depthwise separable convolutions}~\cite{howard2017mobilenets}.

\item \textbf{Specificity-driven regularization for generalization}.

\item \textbf{Small-scale initialization}~\cite{luo2021phase}.

\end{itemize}

As a result, our submitted models collectively occupied the top \textbf{one} position on the BraTS 2025 Task-6 (PED) validation leaderboard, highlighting the effectiveness and consistency of our approach across different tumor subtypes and imaging variations.

\section{Model Architecture}

\

In this section, we begin with the standard nnU-Net~\cite{isensee2021nnu} as the baseline and we propose several targeted modifications aimed at further improving segmentation accuracy and robustness. These include the widened residual encoder with squeeze-and-excitation (SE) attention~\cite{hu2018squeeze} modules, depthwise separable convolutions~\cite{howard2017mobilenets}, regularization, and small-scale weight initialization~\cite{hang2025scalable}.

Our final model architecture is illustrated in Fig.~\ref{advanced U-Net}. It retains the classic U-Net~\cite{ronneberger2015u} encoder-decoder topology, but each stage is enhanced by residual connections and SE attention, and all skip connections are preserved to fuse low- and high-level features:

\begin{figure}[h]
	\centering
	\includegraphics[width=0.9\textwidth]{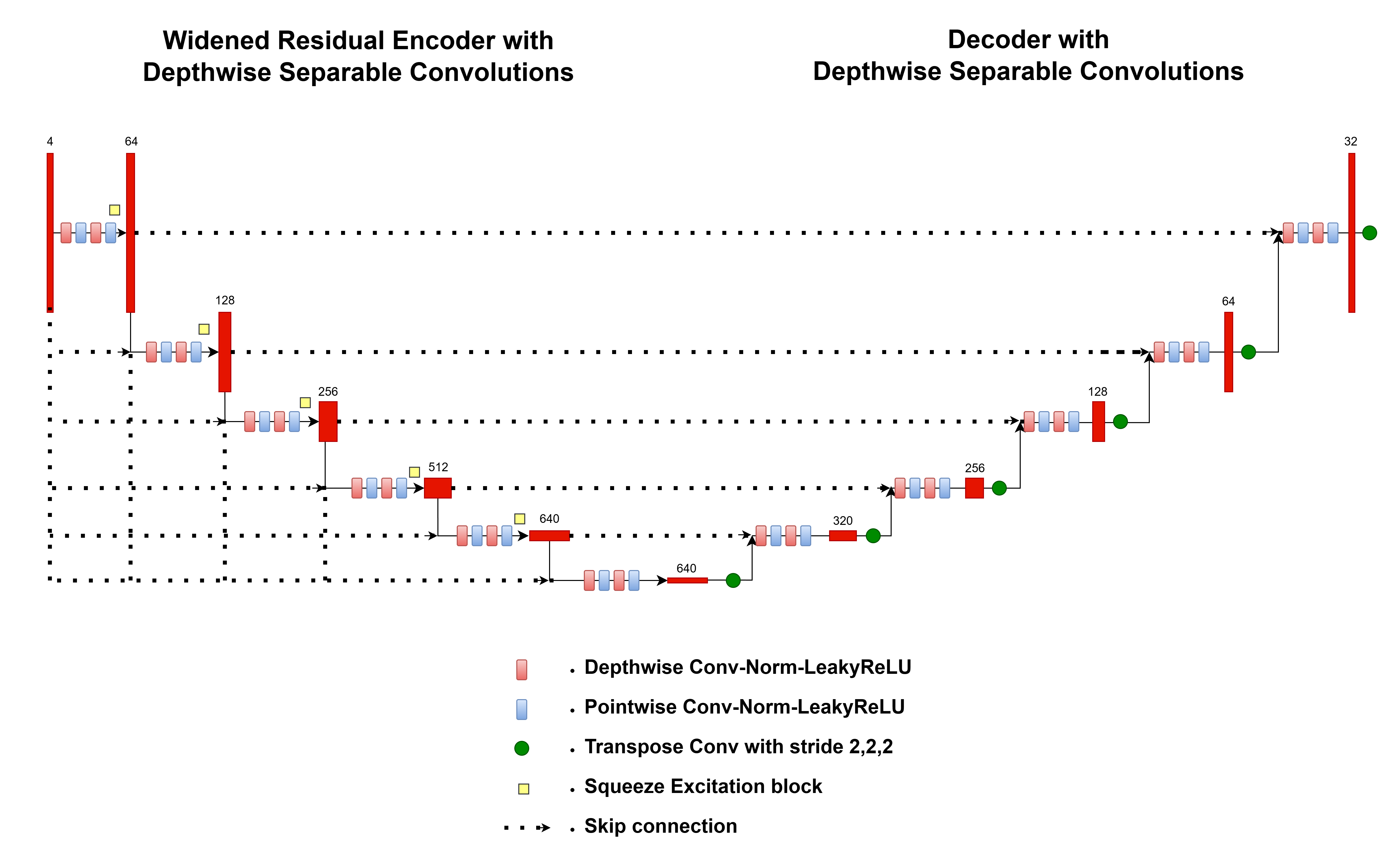}
	\caption{Overview of our enhanced nnU-Net. The left branch is the encoder (downsampling), the right branch is the decoder (upsampling), and dashed arrows denote skip connections.}
	\label{advanced U-Net}
\end{figure}

Downsampling is implemented via a $3\times3\times3$ stride-2 convolution to maintain spatial context. Symmetrically, each upsampling stage begins with a transposed convolution for upscaling.

\subsection{Baseline: standard nnU-Net(v2)}

\

Medical image segmentation is notoriously challenging due to inherent variability across imaging modalities, spatial resolutions, anatomical structures, and pathological features.

To overcome these issues, nnU-Net~\cite{isensee2021nnu} provides a robust, automated pipeline designed specifically for semantic segmentation tasks. Built upon a flexible U-Net architecture, nnU-Net analyzes dataset characteristics, including image dimensionality (2D or 3D), number of modalities, voxel spacings, and class imbalances, automatically generating an optimized configuration without user intervention. This self-adaptation significantly reduces reliance on expert-driven tuning, enabling consistent and high-quality segmentation performance across diverse medical imaging datasets.

For its architecture, nnU-Net by default uses $3\times3\times3$ kernels with strides of 2 (except for the first layer) to replace pooling operations, allowing the model to downsample the feature maps while retaining spatial information. It employs Leaky ReLU activation with a slope of 0.01 to introduce non-linearity and help the model learn more complex representations. Additionally, nnU-Net incorporates Instance Normalization after each convolutional layer, which helps normalize the feature maps and ensures stable training, particularly when handling images with varying intensity distributions.

The practical versatility of nnU-Net has been extensively validated across a wide range of segmentation benchmarks, underscoring its suitability as a reliable baseline in medical image segmentation research. In our work, we adopt the standard nnU-Net (v2) as the baseline, against which we compare our proposed improvements detailed in subsequent sections.

\subsection{Widened residual encoder with SE Attention}

\

To enhance the feature extraction capability of nnU-Net, we first incorporated residual connections into the encoder architecture. Traditional nnU-Net encoders often face challenges such as gradient vanishing and feature degradation when propagating information through multiple convolutional layers. Residual connections effectively alleviate these issues by providing shortcut pathways that facilitate gradient flow and enhance the network's ability to capture complex spatial features.

In addition to residual connections, we further widened the encoder by increasing the number of feature channels in each encoder layer to twice their original values. By widening the encoder, we significantly expanded the model's representational capacity, allowing it to capture richer and more discriminative feature representations. This modification particularly benefits the network's ability to handle intricate structures and subtle variations commonly observed in medical images, ultimately leading to improved segmentation performance and robustness.

\textbf{Residual Blocks with SE Attention}

A single residual block with Squeeze-and-Excitation (SE) attention~\cite{hu2018squeeze} in our encoder is thus defined by

\begin{equation}
\mathbf{y}=\sigma(\mathbf{x}+\mathrm{SE}(\operatorname{DropPath}(\mathcal{F}(\mathbf{x})))),
\end{equation}

where
\begin{itemize}
\item  $\mathcal{F}(\mathbf{x})$ is the stacked convolutional path fitting the residual mapping $\mathcal{H}(\mathbf{x})-\mathbf{x}$.
\item  DropPath $(\cdot)$ applies stochastic depth with drop probability $p=0.05$.
\item  $\mathrm{SE}(\cdot)$ denotes the squeeze excitation attention with reduction ratio $1 / 16$.
\item  $\sigma(\cdot)$ is the final nonlinearity.
\end{itemize}

We find that applying residual blocks solely in the encoder yields the best segmentation accuracy and generalization.

\textbf{Widened Encoder:}

To further boost feature-extraction capacity, we widen the encoder by increasing its channel dimensionality to twice that of the decoder at each corresponding stage. Concretely, if the decoder stages use $\{F_1, F_2, \ldots, F_L\}$ feature maps (e.g. $32,64,128,256,320,320$ ), then the encoder stages are configured with $\{2 F_1, 2 F_2, \ldots, 2 F_L\}$ feature maps (i.e. $64,128,256,512,640,640$ ). This doubling applies to both the initial convolution in each stage and all residual blocks within that stage.

By allocating more channels in the encoder, the network can capture a richer set of spatial and textural features before down-sampling, which in turn allows the decoder (with half the channels) to reconstruct finer details more accurately. Our experiments show that this widened encoder configuration yields a consistent improvement of $2-4 \%$ in overall Dice score on the validation set, as well as better generalization on small and low-contrast tumor regions.

\subsection{Depthwise Separable Convolutions}

\

\begin{figure}[h]
	\centering
	\includegraphics[width=0.6\textwidth]{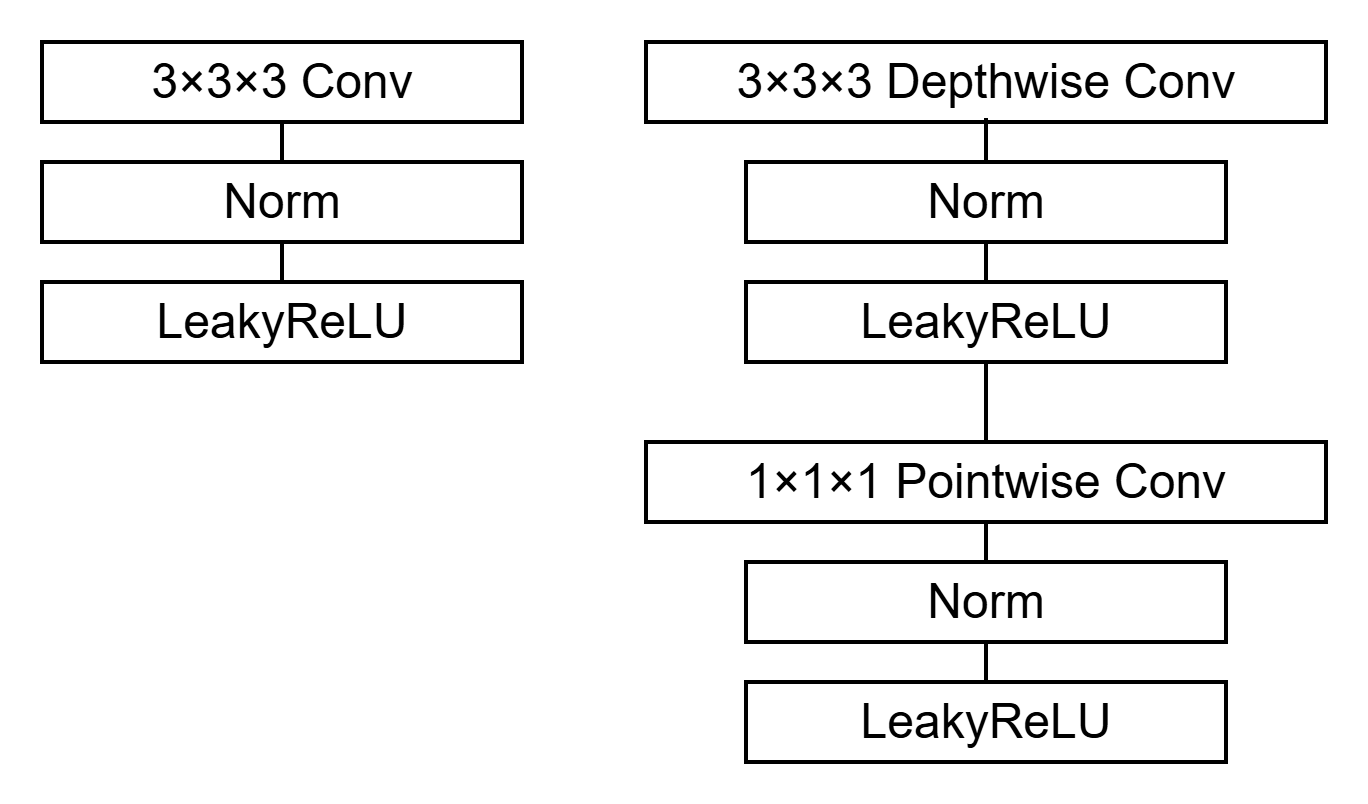}
	\caption{Left: Standard convolution with norm and LeakyReLU. Right: Depthwise Separable convolutions with norm and LeakyReLU.}
	\label{Depthwise Separable Convolutions}
\end{figure}

This section explains the concept of depthwise separable convolution (Fig.~\ref{Depthwise Separable Convolutions} Right) in 3D, including its parameters and computational details.

\subsubsection{Standard Convolution}

\

\begin{figure}[h]
    \centering
    \includegraphics[width=0.25\linewidth]{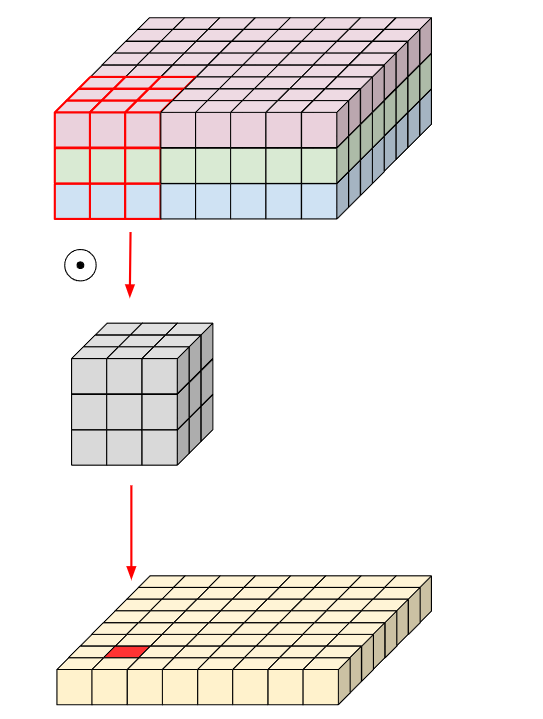}
    \caption{Standard Convolution}
    \label{Standard Convolution}
\end{figure}

\textbf{Standard convolution} (Fig.~\ref{Standard Convolution}~\cite{bendersky2018depthwise}) in 3D applies a kernel to the input to produce the output.

\begin{itemize}
    \item \textbf{Kernel}: A tensor of size $ (k, k, k, C_{in}, C_{out}) $, where $ (k, k, k) $ is the kernel size.
    \item \textbf{Parameter Count}: The total number of parameters is $ k^3 \cdot C_{in} \cdot C_{out} $.
\end{itemize}

\subsubsection{Depthwise Convolution}

\

\begin{figure}[h]
  \centering
  \subfigure[Depthwise Convolution]{%
    \includegraphics[width=0.48\linewidth]{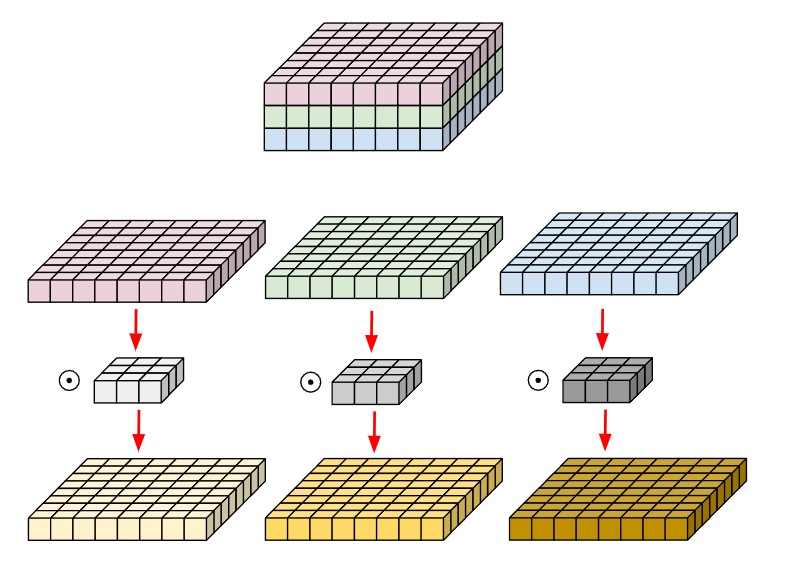}%
    \label{fig:depthwise}%
  }%
  \hfill
  \subfigure[Pointwise Convolution]{%
    \includegraphics[width=0.48\linewidth]{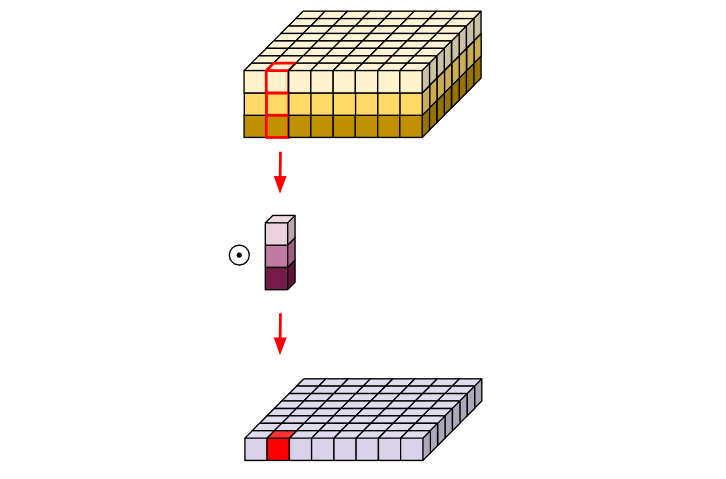}%
    \label{fig:pointwise}%
  }%
  \caption{Depthwise and Pointwise Convolution}
  \label{fig:depth_point}
\end{figure}

\textbf{Depthwise convolution} (Fig.~\ref{fig:depthwise}~\cite{bendersky2018depthwise}) applies a single filter to each input channel independently.

\begin{itemize}
    \item \textbf{Kernel}: A tensor of size $ (k, k, k, 1, C_{in}) $, where $ k $ is the kernel size.
    \item \textbf{Parameter Count}: The total number of parameters is $ k^3 \cdot C_{in} $.
\end{itemize}

\subsubsection{Pointwise Convolution}

\

\textbf{Pointwise convolution} (Fig.~\ref{fig:pointwise}~\cite{bendersky2018depthwise}) uses a $ 1 \times 1 $ kernel to combine the outputs of the depthwise convolution.

\begin{itemize}
    \item \textbf{Kernel}: A tensor of size $ (1, 1, 1, C_{in}, C_{out}) $.
    \item \textbf{Parameter Count}: The total number of parameters is $C_{in} \cdot C_{out} $.
\end{itemize}

\subsubsection{Depthwise Separable Convolution}

\

Depthwise separable convolution~\cite{howard2017mobilenets} consists of two parts: a \textbf{depthwise convolution} and a \textbf{pointwise convolution}. Depthwise convolution integrates information within each channel, while pointwise convolution fuses information across channels. To enhance the model’s expressive capacity, we insert an activation function between the depthwise convolution and the pointwise convolution.

\subsection{Regularization}

\

The evaluation criteria for BraTS2025 introduced lesion-wise Dice as a crucial performance metric. During model training and validation, we observed some cases within the validation set lacking specific lesion classes for example, cases without \textbf{enhancing tumors (ET)}. However, because our neural network model predicts probabilities at the voxel level, it becomes inherently difficult for the model to produce outputs completely absent of certain classes (i.e., predictions that are uniformly zero). This challenge aligns with our understanding of neural network behavior, as such predictions correspond to high-frequency, sparse outputs that networks generally find difficult to accurately learn~\cite{xu2019frequency}.

Under lesion-wise Dice evaluation, predictions containing even minor false positives (FP) for absent classes lead to a Dice score of 0, whereas correctly predicting an absence (output of all zeros) yields a perfect score of 1. This substantial discrepancy significantly impacts overall performance. To address this and improve model accuracy in predicting cases with absent classes, we specifically introduced a regularization term to penalize false positives.

Our proposed approach integrates multiple loss functions covering different segmentation aspects (distribution-based, region-based, and boundary-based) with an additional \textbf{specificity-driven regularization}.

\begin{equation}
\text{Loss} = -\text{Dice}+\text{CE}+\text{HD}+\frac{\theta N_\text{FP}}{N_\text{pred}+N_\text{gt}}
\end{equation}

We set $\theta = 0.1$. By explicitly penalizing false positives through this regularization term, our model demonstrates enhanced predictive accuracy for cases lacking specific lesion classes, thereby significantly improving the lesion-wise Dice performance.

\subsection{Initialization}

\

The initialization of neural networks across different scales significantly influences their generalization capability~\cite{luo2021phase}. This effect is closely related to phenomena such as condensation and the network’s Hessian eigenvalues~\cite{li2023loss}, which determine how a network converges during training. The impact of initialization is not only observable in simpler architectures like fully connected neural networks, but it extends to more complex models such as Convolutional Neural Networks (CNNs), ResNet~\cite{he2016deep}, and even large language models. Initialization plays a crucial role in the dynamics of training and the network's ability to generalize to unseen data.

Generally speaking, smaller initialization~\cite{luo2021phase} values tend to favor the network's reasoning ability rather than its memory capacity~\cite{hang2025scalable}. This characteristic is particularly beneficial for tasks that require strong reasoning capabilities. In such tasks, models initialized with smaller values are typically more effective at capturing general patterns rather than memorizing specific details.

We utilize Gaussian initialization, which has been shown to yield good performance in a wide range of deep learning architectures. Mathematically, Gaussian initialization is typically expressed as follows:

\begin{equation}
w \sim \mathcal{N}\left(0, (\frac{2}{dim_{in}})^{\alpha}\right)
\end{equation}

where $w$ represents the weights of the network, and $n_{i n}$ represents the input feature dimension of the convolutional layer, and $\alpha$ is a hyperparameter introduced to control the scale of initialization. By tuning the value of $\alpha$, we effectively control the scale of initial weights.

\subsection{Postprocessing}

\

We primarily implemented two postprocessing techniques to further refine the segmentation results. 

The first technique leverages domain-specific medical imaging knowledge, particularly focusing on \textbf{enhancing tumor (ET)} segmentation accuracy. Given that ET is a critical region in tumor identification and typically occupies smaller volumes compared to \textbf{non-enhancing tumors (NET)}, neural networks often struggle to accurately detect \textbf{ET} due to the limited representation and subtle intensity differences. However, exploiting the intensity contrast between T1CE (contrast-enhanced T1-weighted) and T1 (non-enhanced T1-weighted) modalities in MRI scans provides valuable information to better distinguish ET from NET.

From fundamental medical imaging principles, it is known that the ratio of T1CE to T1 signal intensities can effectively differentiate enhancing from non-enhancing tumor regions. To systematically apply this knowledge, we first performed z-score normalization on both T1CE and T1 signals within the training dataset. Subsequently, we calculated the T1CE/T1 intensity ratio specifically at locations annotated as label 1 \textbf{(ET)} and label 2 \textbf{(NET)}. To ensure robustness and avoid outlier influence, we excluded extreme values (ratios below 0.2 and above 5) from our analysis.

\begin{figure}[h]
    \centering
    \includegraphics[width=0.7\linewidth]{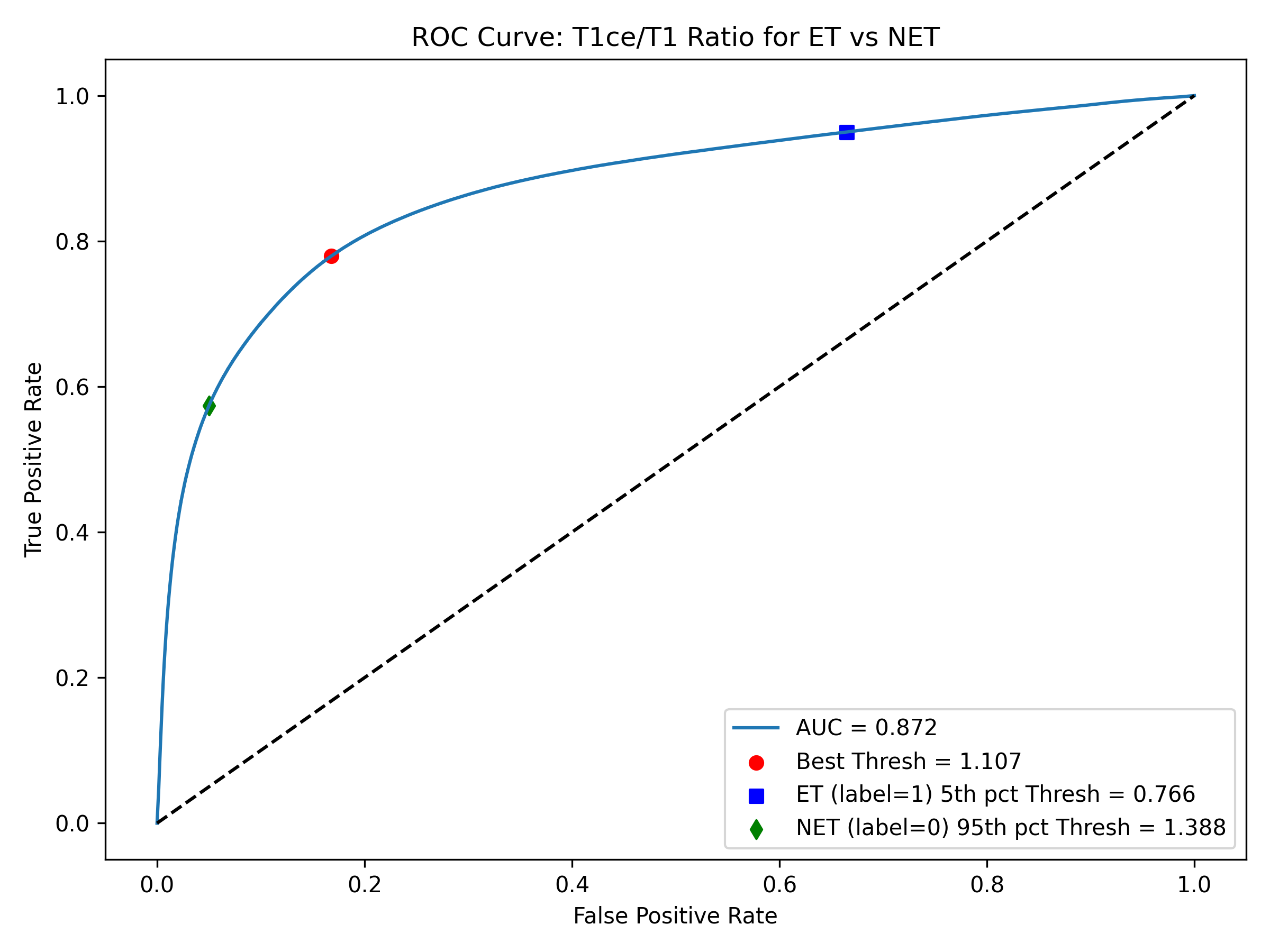}
    \caption{ROC Curve of T1CE-T1 ratio}
    \label{ROC}
\end{figure}

Based on the statistical analysis, we conservatively selected the 95-th percentile values as threshold criteria: specifically, we reassigned voxels initially labeled as \textbf{NET} (label 2) to \textbf{ET} (label 1) if their T1CE/T1 ratio exceeded 1.388. Conversely, voxels initially labeled as \textbf{ET} (label 1) were reassigned to \textbf{NET} (label 2) if their ratio fell below 0.766. Our ROC curve is show in Fig.~\ref{ROC}.

The second postprocessing approach focuses on removing small isolated connected components.  First, we apply a $3\times3\times3$ dilation kernel to the voxel-wise predictions. Afterward, connected components are identified, and their volumes are calculated. Through threshold testing on the validation dataset, we determined optimal volume thresholds of $160 mm^3$ and $50 mm^3$ for labels 1 and 3, respectively. This approach effectively reduces false-positive predictions by removing small connected components and enhances the spatial consistency of the segmented structures.

\section{Training}

\

We train our network using 5-fold cross-validation with the SGD optimizer with weightdecay = 3e-5 and momentum = 0.99 for $1000$ epochs and a batch size of $2$ on NVIDIA GeForce RTX 4080 16GB GPUs. In each epoch we randomly sample 250 patches; the initial learning rate is set to 1e-2. We decay the learning rate according to a cosine schedule over the full 1000 epochs:

\begin{equation}
\eta_t=\eta_0 \frac{1+\cos (\pi t / T)}{2}
\end{equation}

\begin{figure}[h]
  \centering
  \subfigure[Loss and Dice]{%
    \includegraphics[width=0.48\linewidth]{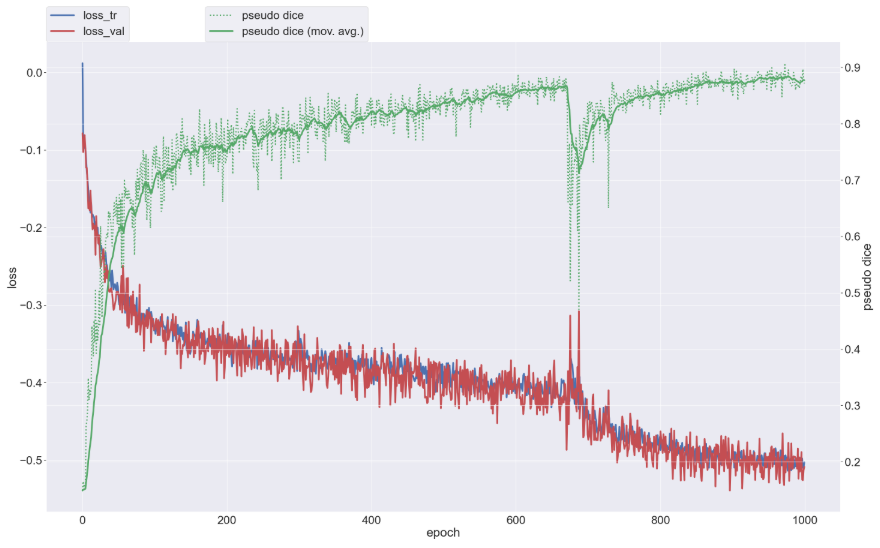}%
    \label{fig:loss_dice}%
  }\hfill
  \subfigure[Learning Rate]{%
    \includegraphics[width=0.48\linewidth]{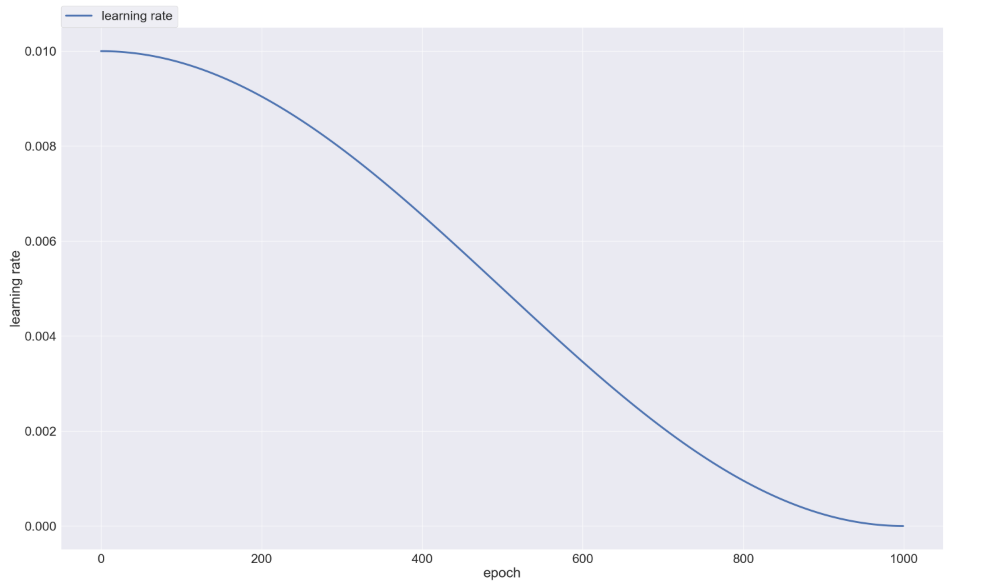}%
    \label{fig:lr_schedule}%
  }%
  \caption{(a) Training loss and Dice score curves. \quad (b) Cosine learning‑rate schedule over epochs.}
  \label{train}
\end{figure}

Our training data augmentations are as follows: we apply spatial transforms (elastic deformation, random rotations, scaling); add Gaussian noise and Gaussian blur; perform multiplicative brightness and contrast adjustments; simulate low-resolution sampling; apply two gamma corrections; and randomly flip along all three axes. Fig.~\ref{train} illustrates the evolution of the training loss (Fig.~\ref{fig:loss_dice}), Dice score (Fig.~\ref{fig:loss_dice}), and learning rate (Fig.~\ref{fig:lr_schedule}) over the course of training.

\section{Results}

\

On the BraTS-PED (Task 6) validation leaderboard, our method ranks first with the following LesionWise Dice scores:

\begin{table}[h!]
\centering
\caption{\textbf{Lesion-wise Dice results on validation dataset.} Higher values indicate better performance.}
\renewcommand{\arraystretch}{1.2}
\setlength{\tabcolsep}{5pt}
\resizebox{0.8\linewidth}{!}{%
\begin{tabular}{lcccccc}
\toprule
\textbf{} & \textbf{CC} & \textbf{ED} & \textbf{ET} & \textbf{NET} & \textbf{TC} & \textbf{WT} \\
\midrule
\textbf{value} & 0.759 & 0.967 & 0.826 & 0.910 & 0.928 & 0.928 \\
\bottomrule
\end{tabular}%
}
\end{table}

\begin{figure}[h]
  \centering
  \subfigure[Sagittal of case 381.]{%
    \includegraphics[width=0.33\linewidth]{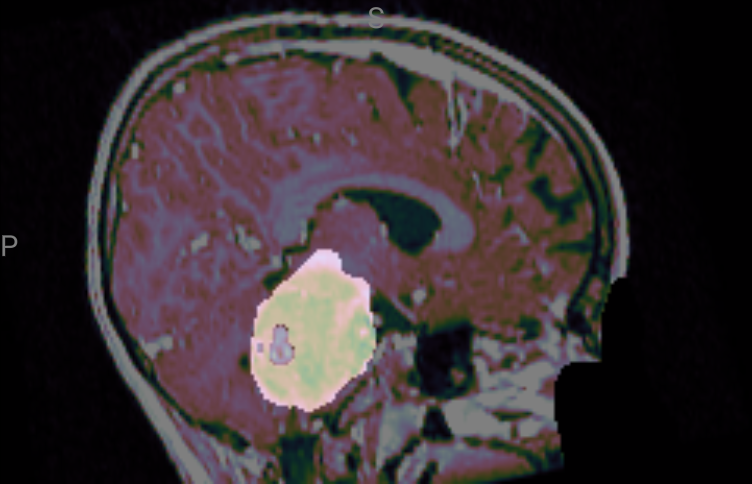}%
    \label{fig:case381_sagittal}%
  }\hfill
  \subfigure[Coronal of case 381.]{%
    \includegraphics[width=0.33\linewidth]{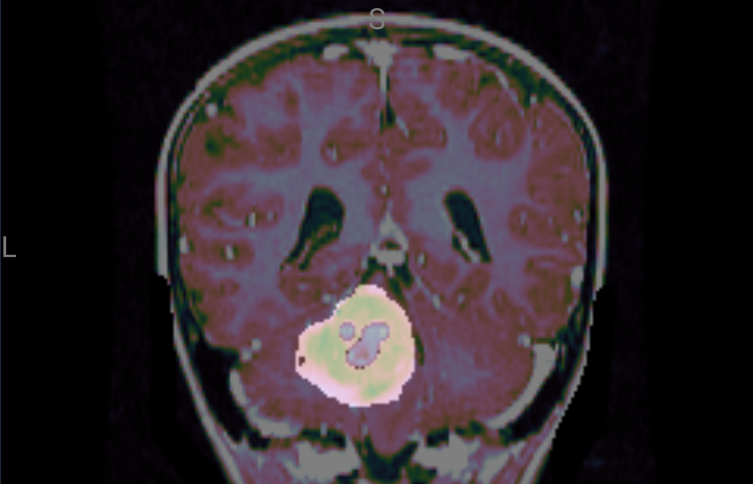}%
    \label{fig:case381_coronal}%
  }\hfill
  \subfigure[Axial of case 381.]{%
    \includegraphics[width=0.3\linewidth]{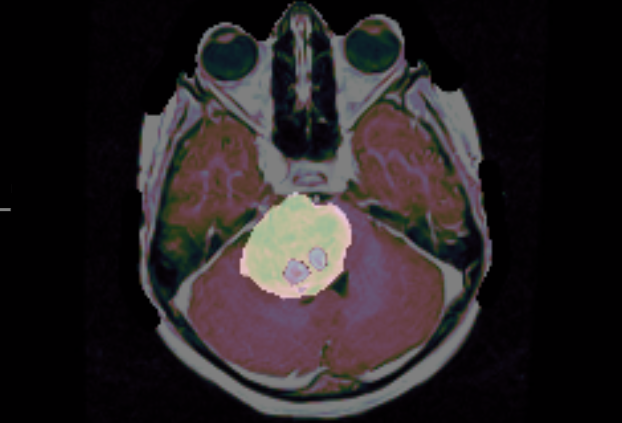}%
    \label{fig:case381_axial}%
  }%
  \caption{BraTS\_PED\_00381\\
    ET: 0.9575\quad NET: 0.9722\quad TC: 0.9749\quad WT: 0.9749}
  \label{Case381}
\end{figure}

These results demonstrate great performance across all critical tumor cystic component (CC), peritumoral edema (ED), enhancing tumor (ET), non-enhancing tumor (NET), tumor core (TC), and whole tumor (WT). In Fig.~\ref{Case381}, we present a representative example in which our model delivers highly accurate lesion segmentation, clearly illustrating its precise predictive capabilities.

On the BraTS-PED (Task 6) \textbf{test set}, our method achieves excellent performance, ranking among the top results with the following quantitative metrics:

\begin{table}[h!]
\centering
\caption{\textbf{Lesion-wise Dice results on test dataset.} 
Higher values indicate better performance.}
\renewcommand{\arraystretch}{1.2}
\setlength{\tabcolsep}{3.5pt} 
\resizebox{0.9\linewidth}{!}{%
\begin{tabular}{lcccccc|cccccc}
\toprule
\textbf{} &
\multicolumn{6}{c}{\textbf{Lesion-wise Dice} $\uparrow$} &
\multicolumn{6}{c}{\textbf{Lesion-wise NSD-1.0} $\uparrow$} \\
\cmidrule(lr){2-7} \cmidrule(lr){8-13}
 & CC & ED & ET & NETC & TC & WT & CC & ED & ET & NETC & TC & WT \\
\midrule
\textbf{mean} 
& 0.591 & \textbf{0.892} & \textbf{0.727} & \textbf{0.838} & \textbf{0.903} & \textbf{0.900}
& 0.599 & \textbf{0.892} & \textbf{0.783} & \textbf{0.800} & \textbf{0.775} & \textbf{0.770} \\
std 
& 0.464 & 0.312 & 0.307 & 0.211 & 0.141 & 0.143 
& 0.459 & 0.312 & 0.297 & 0.213 & 0.220 & 0.230 \\
\bottomrule
\end{tabular}%
}
\end{table}

\section{Discussion}

\

Overall, our model achieves state-of-the-art lesion-wise performance through a combination of architectural innovations, enhanced learning strategies, careful initialization, and task-specific post-processing. Together, these modifications enable richer spatial and contextual encoding of tumor subregions, contributing to our high lesion-wise Dice scores across all targets.

Despite these advances, there remains substantial room for improvement in the ET and CC metrics, especially in reducing false positives. Furthermore, while convolutional neural networks continue to dominate in medical image segmentation, recent fully-Transformer architectures~\cite{wald2025primus} have demonstrated strong performance on 3D medical image segmentation tasks. The relative underperformance of transformer models here likely stems from limited training data to exploit their full representation power and from our preliminary exploration of such designs. Future work should therefore search deeper into attention mechanisms.

\subsubsection{\discintname}
\small 
 The authors have no competing interests to declare that are relevant to the content of this article.

%
%
%
\bibliographystyle{splncs04}
\bibliography{bibfile}





\end{document}